\documentclass[preprintnumbers,amsmath,amssymb,showpacs]{revtex4}

\usepackage{graphicx}
\usepackage{dcolumn}
\usepackage{bm}
\usepackage{graphicx}
\usepackage{booktabs}
\usepackage{CJK}
\usepackage{amsmath}
\usepackage{amssymb}
\usepackage{color} 

\begin{document}

\title{Scalable  Bell inequalities for multiqubit systems}
\author{Ying-Qiu He$^{1}$, Dong Ding$^{1,2}$, Feng-Li Yan$^{1}$ }
\email{flyan@hebtu.edu.cn}
\author{Ting Gao$^3$}
\email{gaoting@hebtu.edu.cn}
\affiliation {$^1$ College of Physics Science and Information Engineering, Hebei Normal University, Shijiazhuang 050024, China \\
$^2$Department of Basic Curriculum, North China Institute of Science and Technology, Beijing 101601, China \\
$^3$College of Mathematics and Information Science, Hebei Normal University, Shijiazhuang 050024, China}
\date{\today}

\begin{abstract}
Based on Clauser-Horner-Shimony-Holt inequality, we show a fruitful method to exploit Bell inequalities for multipartite qubit systems.
These Bell inequalities are designed with a simpler architecture tailored to experimental demonstration.
Under the optimal setting we derive a set of compact Mermin-type inequalities and discuss quantum violations for generalized Greenberger-Horne-Zeilinger (GGHZ) states. Also, we reveal relationship between quantum nonlocality and four-partite entanglement for four-qubit GGHZ states.

\end{abstract}

\pacs{03.65.Ud, 03.67.Mn}

\maketitle

\section{Introduction}
In 1964, the first Bell inequality was exploited  \cite{Bell1964}. Based on local hidden variable model (LHVM),
Bell inequalities can always give upper bounds on certain quantities. In quantum mechanics, if a higher value is obtained
than these bounds, it is commonly referred to as a quantum violation.
Although it has unsatisfactory aspect of applying only to a statistical measurement procedure,
the Bell inequalities are still of great importance in investigating quantum entanglement \cite{HHHH2009, GT2009, DVC2000, YGC2011, GYE2014}.

Inspired by Bell's paper, then, Clauser, Horne, Shimony, and Holt (CHSH) \cite{CHSH1969} derived a correlation inequality, which involves a bipartite correlation function (average over many runs of experiment) for two alternative dichotomic observables. It has been pointed out that \cite{Gisin1991, GP1992} any pure entangled state of two particles violates the correlation CHSH inequality. Later, the studies of Bell inequalities are expanded to three-body systems \cite{Svetlichny1987,ACGKKOZ2004}.
For $n$-partite quantum system, there are several fruitful Bell inequalities such as Mermin-Ardehali-Belinski-Klyshko (MABK) inequalities \cite{Mermin1990, Ardehali1992, BK1993}, or the more general Werner-Wolf-\.{Z}ukowski-Brukner (WWZB) inequalities \cite{WW2001, ZB2002}, and other type inequalities \cite{SS2002, LF2012, WZCG2013}.
By far, Bell inequalities are an important tool to investigate the possible connection between nonlocality and entanglement \cite{GSDRS2009, AR2010, GDSKS2010, DYG2013-10}. However, with the increase of the parties more and more items ($2^n$ for WWZB inequalities, for example) will be a drawback in practice and it is still an important task to seek some optimal Bell inequalities.

In this paper, we present a method to construct the scalable   Bell inequalities for multiqubit systems. Based on CHSH inequality, we establish a set of
Bell inequalities, where the number of terms grows exponentially with half of number of particles.
In the structure of the present inequalities, the standard CHSH Bell inequality can be obtained for $n=2$ and, as an example, compact Mermin-type inequalities are proposed.

\section{Bell inequalities}

Consider an $n$-qubit system, where each of the parties is allowed to choose independently between two dichotomic observables $A^i_1$, $A^i_2$ for the $i$-th observer, $i=1,2,...,n$, and each outcome can either take value $+1$ or $-1$. Now, we first review a well known Bell-type inequality for bipartite systems, CHSH inequality \cite{CHSH1969},
\begin{equation}\label{}
|\langle A^1_1A^2_1+A^1_1A^2_2+A^1_2A^2_1-A^1_2A^2_2 \rangle|\leq2.
\end{equation}
 Consider two partitions of CHSH polynomial $A^1_1A^2_1-A^1_2A^2_2$ and $A^1_1A^2_2+A^1_2A^2_1$. Obviously, based on absolute LHVM,
there exist two possible outcomes: $A^1_1A^2_1-A^1_2A^2_2=0$ and $A^1_1A^2_2+A^1_2A^2_1=\pm 2$, or $A^1_1A^2_1-A^1_2A^2_2=\pm2$ and $A^1_1A^2_2+A^1_2A^2_1=0$.
In terms of the particular relevance, then, we provide an efficient algorithm for generating the scalable  Bell inequalities for multiqubit systems.

In this architecture, elementary units of the present inequalities are paired partitions of CHSH polynomial. More specifically,
when $n$ is even the polynomials read
\begin{equation}\label{}
\mathcal{B}=\frac{1}{\sqrt{2^{n}}}[\prod_{j~\mathrm{is~odd}}(A^j_1A^{j+1}_1-A^j_2A^{j+1}_2)+(-1)^{\lfloor \frac{n-1}{4}\rfloor} \prod_{j~\mathrm{is~odd}}(A^j_1A^{j+1}_2+A^j_2A^{j+1}_1)],
\end{equation}
for odd $n$  we get
\begin{equation}\label{}
\mathcal{B}=\frac{1}{\sqrt{2^{n-1}}}[\prod_{j~\mathrm{is~even}}A^1_1(A^j_1A^{j+1}_1-A^j_2A^{j+1}_2)+(-1)^{\lfloor \frac{n-1}{4}\rfloor} \prod_{j~\mathrm{is~even}}A^1_2(A^j_1A^{j+1}_2+A^j_2A^{j+1}_1)].
\end{equation}
Actually, for  odd $n$ , the products can be viewed as considerable generalization by multiplying a pair of trivial polynomials $A^i_1$ and $A^i_2$ on any one of the observers $i$, letting $i=1$, for example.
From the relationship between two partitions $A^j_1A^{j+1}_1-A^j_2A^{j+1}_2$ and $A^j_1A^{j+1}_2+A^j_2A^{j+1}_1$,
algebraically, we have $| \mathcal{B} |\leq 1$.
In a local classical model, letting $\langle\mathcal{B}\rangle$ denote mean value of $\mathcal{B}$, one can show that
inequalities $|\langle\mathcal{B}\rangle|\leq 1$ holds.

\section{Quantum violations}
We now investigate quantum violations of the present Bell inequalities. As quantum counterpart of the Bell polynomial, Bell operator $\mathcal{B}$ should satisfy inequality
\begin{equation}\label{Bleq1}
\mathrm{tr}(\varrho \mathcal{B})=|\langle\mathcal{B}\rangle|\leq 1,
\end{equation}
where density operator $\varrho$ is used to character an $n$-partite quantum system. A quantum violation of one of the inequalities refer to the result of the left hand side of expression (\ref{Bleq1}) larger than 1.
In generally, quantum mechanically, one expresses the observables as $A^i_1=\vec{a}^i_1\cdot \vec{\sigma}$, $A^i_2=\vec{a}^i_2\cdot \vec{\sigma}$, $i=1,2,...,n$, where $\vec{a}^i_1$, $\vec{a}^i_2$ are unit vectors and $\vec{\sigma}=(\sigma_x, \sigma_y, \sigma_z)$ denotes a vector of Pauli matrices.

Consider $n$-qubit Greenberger-Horne-Zeilinger (GHZ) state
\begin{equation}\label{GHZ}
     \left|\psi_n\right\rangle=
     \frac{1}{\sqrt{2}}(\left| {0} \right\rangle^{\otimes n}  + \left| {1} \right\rangle^{\otimes n}).
\end{equation}
Notice that it is an eigenstate of operator
\begin{equation}\label{nsigma}
\mathcal{A}_n=\frac{1}{2}(\prod^{n}_{j=1}(\sigma_x+\texttt{i}\sigma_y)_j+\prod^{n}_{j=1}(\sigma_x-\texttt{i}\sigma_y)_j).
\end{equation}
with eigenvalue $2^{n-1}$. As is well known, GHZ states maximally violate Mermin inequalities \cite{Mermin1990}. The Mermin inequalities have $2^{n-1}$ terms and bound the values of quantum violation $2^{(n-1)/2}$ for odd $n$, and $2^{n/2-1}$ for even $n$. It is worth noting that the present inequalities involve a series of compact Mermin-type inequalities.
In detail, for $n=4k+2$, $k=0,1,2,\cdots$, taking the experimental setting $\vec{a}^1_1=({1}/{\sqrt{2}},{1}/{\sqrt{2}},0)$, $\vec{a}^1_2=(-{1}/{\sqrt{2}},{1}/{\sqrt{2}},0)$, $\vec{a}^i_1=(1,0,0)$, $\vec{a}^i_2=(0,1,0)$, $i=2,3,\cdots,n$, we have $\mathrm{tr}[\varrho (|\psi_n\rangle)\mathcal{B}]=\sqrt{2}$.
For $n=4k+3$ and $n=4k+4$, we take $\vec{a}^i_1=(1,0,0)$, $\vec{a}^i_2=(0,1,0)$, $i=1,2,\cdots,n$ and obtain $\mathrm{tr}[\varrho (|\psi_n\rangle)\mathcal{B}]=2$.
For $n=4k+5$, taking $\vec{a}^1_1=\vec{a}^1_2=(1,0,0)$, $\vec{a}^i_1=(1,0,0)$, $\vec{a}^i_2=(0,1,0)$, $i=2,3,\cdots,n$, also, we get $\mathrm{tr}[\varrho (|\psi_n\rangle)\mathcal{B}]=2$.
In a word, under the optimal setting we show that the compact Mermin-type inequalities, which have $2^{(n+1)/2}$ terms for odd $n$ and $2^{n/2+1}$ for even $n$,  are maximally violated by $n$-qubit GHZ states with a certain constant visibility $\sqrt{2}$ or 2.

As an example, we now use the present inequalities to test the quantum nonlocality of $n$-qubit generalized GHZ (GGHZ) states
\begin{equation}\label{GGHZ}
  \left|\psi_n(\alpha)\right\rangle=\cos{\alpha}\left| {0} \right\rangle^{\otimes n} + \sin{\alpha}\left| {1} \right\rangle^{\otimes n}.
\end{equation}
Under the specified setting, as already mentioned, one can obtain
\begin{equation}\label{}
\mathrm{tr}[\varrho (|\psi_n(\alpha)\rangle)\mathcal{B}]=\sqrt{2}\sin{2\alpha}
\end{equation}
for $n=4k+2$, and
\begin{equation}\label{2sinalpha}
\mathrm{tr}[\varrho (|\psi_n(\alpha)\rangle)\mathcal{B}]=2\sin{2\alpha}
\end{equation}
for the others. That is, for $\sin{2\alpha} > {1}/{2}$ (or $\sin{2\alpha} > {1}/{\sqrt{2}}$ for $n=4k+2$) the present inequalities are violated. Compared with the threshold $\sin{2\alpha}={1}/{\sqrt{2^{n-1}}}$ described in \cite{SG2001}, we can find that the decrease of items of correlation functions is sometimes at the cost of detecting precision.

\begin{figure}
  \centering\includegraphics[width=3.5in]{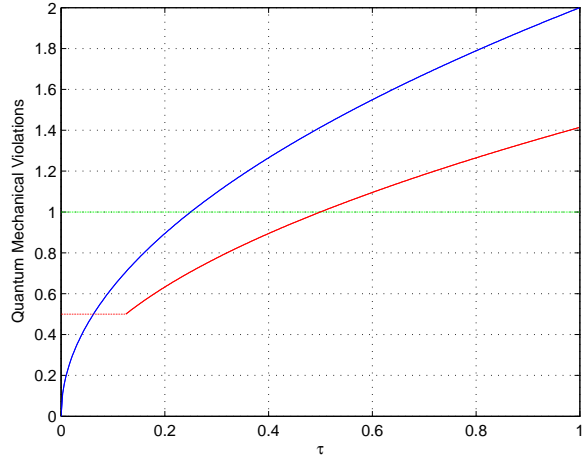}\\
  \caption{(color online). Numerical results of quantum nonlocality versus four-partite entanglement.}\label{picture-B-tau}
\end{figure}

Next, we analyze the relationship between quantum nonlocality and entanglement for GGHZ states. For simplicity, we here focus on the situation of $n=4$.
Particularly, we recast the four-qubit Bell polynomial as
\begin{equation}\label{B-sigma}
\mathcal{B}=\frac{1}{4}(\sigma_x\sigma_x\sigma_x\sigma_x +\sigma_y\sigma_y\sigma_y\sigma_y-\sum_\mathrm{perm}\sigma_x\sigma_x\sigma_y\sigma_y)
\end{equation}
and thus the expression (\ref{2sinalpha}) holds.
On the other hand, consider the four-tangle of four-qubit GGHZ state
\begin{equation}\label{}
\tau(|\psi_4(\alpha)\rangle)=\sin^2{2\alpha},
\end{equation}
where the four-tangle $\tau$ quantifies four-partite entanglement \cite{Four-tangle2010}.
Then, we can recover the entanglement-nonlocality relationship
\begin{equation}\label{equation-B-tau}
\mathrm{tr}[\varrho (|\psi_4\rangle)\mathcal{B}]=2\sqrt{\tau}.
\end{equation}
By the way, we also note that, for generalized slice-state \cite{CS2000}
\begin{equation}\label{}
  \left|\psi_\mathrm{s}\right\rangle=\cos{\alpha}\left| {0000} \right\rangle  + \sin{\alpha}\left| {1} \right\rangle(\cos{\beta}\left| {0} \right\rangle  + \sin{\beta}\left| {1} \right\rangle)(\cos{\gamma}\left| {0} \right\rangle  + \sin{\gamma}\left| {1} \right\rangle)(\cos{\delta}\left| {0} \right\rangle  + \sin{\delta}\left| {1} \right\rangle),
\end{equation}
the relationship (\ref{equation-B-tau}) hold.
In Fig.\ref{picture-B-tau}, we show the numerical results of quantum nonlocality versus entanglement for the four-qubit GGHZ states.
The green line is the threshold above which the inequalities are violated. It has been shown that (see blue line), for $\tau>1/4$, the quantum violation of the present inequality occurs and varies with four-tangle. Also, we plot a relationship between entanglement and the maximum expectation value of a generalized Svetlichny operator for four-qubit GGHZ states \cite{GDSKS2010} denoted by the red line, where the generalized Svetlichny inequality is only violated with $\tau>1/2$. Therefore, in a sense, the present inequality is more sensitive to detecting the four-qubit GGHZ states.

\section{Conclusion}

In conclusion, we have shown a fruitful method for constructing a series of correlation Bell inequalities for multipartite qubit systems.
In the architectures of these Bell inequalities, paired partitions of CHSH inequality are exploited.
Under the optimal setting we have reported a set of compact Mermin-type inequalities and discussed the maximal quantum violations. Using the present inequalities, we have studied quantum nonlocality of $n$-qubit GGHZ states. Also, we have obtained a useful relationship between quantum nonlocality and four-partite entanglement for four-qubit GGHZ states.
Moreover, it is worth pointing out that the total numbers of correlation functions of the present inequalities have been greatly reduced, growing exponentially with half of number of qubits. Together with there are only two dichotomic observables per site, this make it convenient to design a scalable scheme for testing the quantum formalism against the LHVM in experiments.

\section*{Acknowledgements}
This work was supported by the National Natural Science Foundation of China under Grant Nos: 11475054,11371005, Hebei Natural Science Foundation of China under Grant Nos: A2012205013, A2014205060,  the Fundamental Research Funds for the Central Universities of Ministry of Education of China under Grant No: 3142014068, Langfang Key Technology Research and Development Program of China under Grant No: 2014011002.

\end{document}